\documentclass[preprint,showpacs,superscriptaddress,floatfix,prl]{revtex4-1}

\usepackage{amsmath}
\usepackage[dvips]{graphicx}
\usepackage{epsfig}
\usepackage{tabularx}
\newcommand{\ud}{\mathrm{d}}

\begin{document}

\title{Photothermal and thermo-refractive effects in high reflectivity mirrors at room and cryogenic temperature}

\author{Alessandro~Farsi, Mario~Siciliani~de~Cumis, Francesco~Marino and Francesco~Marin}

\affiliation{Dipartimento di Fisica, Universit\`a di Firenze,
INFN, Sezione di Firenze, and LENS \\ Via Sansone 1, I-50019 Sesto
Fiorentino (FI), Italy}

\email[Corresponding author: ]{marin@fi.infn.it}

\date{\today}
\begin{abstract}
Increasing requirements in the sensitivity of interferometric measurements is a common feature of several research fields, from gravitational wave detection to quantum optics. This motivates refined studies of high reflectivity mirrors and of noise sources that are tightly related to their structure. In this work we present an experimental characterization of photothermal and thermo-refractive effects in high reflectivity mirrors, i.e., of the variations in the position of their effective reflection plane due to weak residual power absorption. The measurements are performed by modulating the impinging power in the range 10 Hz $\div$ 100 kHz. The experimental results are compared with an expressly derived theoretical model in order to fully understand the phenomena and exploit them to extract useful effective thermo-mechanical parameters of the coating. The measurements are extended at cryogenic temperature, where most high sensitivity experiments are performed (or planned in future versions) and where characterizations of dielectric film coatings are still poor.

\end{abstract}


\maketitle

\section{Introduction}

Extremely sensitive interferometric measurements are implied in several frontier experiments. For instance, large baseline gravitational wave detectors presently surpass the limit of $10^{-22}$~ Hz$^{-1/2}$ in relative displacement spectral noise~\cite{interferometri}. An improvement of more than one order of magnitude is planned for their advanced versions~\cite{interferometri}, and a sensitivity of $10^{-23}$~m/$\sqrt{\textrm{Hz}}$ is envisaged in the kHz range for possible future massive, cryogenic detectors~\cite{dual}. On the small scale, very low-noise background is also obtained in small scale quantum opto-mechanics experiments~\cite{caniard}. 

With such extreme requirements, critical limits are imposed by several noise sources associated with the interferometer optical components~\cite{Brag1}, and a crucial role is played in particular by the high-reflection coating. For instance, Brownian thermal noise of the dielectric layers is the most stringent foreseen limiting factor in the $\sim$kHz range for the next generation of gravitational wave detectors. Another important phenomenon is the photothermal dynamic effect: an (even small) fraction of the light impinging on a mirror is absorbed and heats it, causing a local temperature change and, through thermal expansion, a deformation of the mirror surface. A fundamental limit to the interferometer sensitivity is therefore produced by the shot noise of the absorbed radiation, and such limit can be worse in the case of excess of laser-power fluctuations. This effect, together with other dissipative phenomena such as photoelastic and Brownian noise, has been studied theoretically by Braginsky \emph{et al.} \cite{Brag1} and Cerdonio \emph{et al.} \cite{Cerdonio1} in the case of a Gaussian beam impinging on a half-infinite homogeneous mirror.

In spite of being an annoying noise source, photothermal dynamic displacement can be exploited, using controlled laser power variations, for measuring the thermo-mechanical parameters of the mirror. This possibility is particularly important to investigate the characteristics of the coating. Thin film parameters are generally different from those of the bulk. Moreover, they depend on the film morphology, and finally on the techniques used for deposition. A photothermal analysis gives therefore useful information for understanding the general behavior of the coating and selecting the optimal fabrication procedure. At this purpose, it is important to analyze in particular the low-losses coating and substrates of the kind used in interferometers.

Dynamic photothermal effects were observed in a high-Finesse Fabry-Perot cavity, probably for the first time, by An \emph{et al.}\cite{An}. They describe an hysteretic behavior in the cavity transmission when the laser is rapidly scanned, and use it to estimate the mirror absorption. A more detailed spectral analysis exploiting an amplitude modulated laser source is reported by De Rosa \emph{et al.}\cite{DeRosa02}, who verify the theoretical prediction of Refs. \cite{Brag1} and \cite{Cerdonio1} concerning the photothermal expansion of the mirror substrate.

The study is extended in Ref.~\cite{DeRosa06}, where on one hand the dependence of the photothermal response on the beam waist, foreseen in Ref.~\cite{Cerdonio1}, is verified; on the other side, a deviation from the above theoretical expression is observed at low frequencies ($< 10^{-2}$~Hz) due to the limited mirror size, and at high frequencies ($> 10^2$~Hz) due to the coating.
The effect of the coating is also shown by Black \emph{et al.}\cite{Black}, who apply a gold film on the dielectric coated mirror to enhance the absorption at the expenses of the cavity Finesse. Their dielectric film is composed of alternate layers of SiO$_2$ and TiO$_2$, and their results are in agreement with a simple heuristic model approximating the coating as an homogeneous layer.

In view of their possible use in the future gravitational wave detectors, it is important to check the photothermal effect also on substrates of different materials (particularly interesting are sapphire and silicon) and at cryogenic temperature. Such conditions, explored by the pioneering work of the TAMA group\cite{TAMA}, would greatly improve the sensitivity (at the expenses of evident technical complications).

In the present work we describe a detailed study of photothermal dynamic effect in a high-Finesse cavity, where the mirror coating is very close to those conceived for gravitational wave detectors. Our attention is indeed particularly focused on the behavior of the dielectric films, but we also check the effect on a silicon substrate. Our investigation is performed both at room and at cryogenic temperature, and a theoretical model is developed in order to extract the most reliable information from the experimental data.

\section{Model}
\label{model}

The photothermal expansion is ruled by the thermal diffusion equation and the equation of elasticity. The mirror surface displacement is then read with a weighted average that, in the case of a Gaussian beam, can be written as
\begin{equation}
\bar{u}(t)=\int \ud^2 \mathbf{r}_{\perp}\,u_z(\textbf{r}_{\perp},t)\frac{e^{-\frac{r_{\perp}^2}{r_0^2}}}{\pi r_0^2}
\label{ut}
\end{equation}
where $u_z(\textbf{r},t)$ is the surface normal displacement, $\textbf{r}$ are the transverse coordinates of the mirror surface, $r_0$ is the beam radius measured in correspondence of a power drop by $1/e$ from its central value.

The solution of the problem in the case of an half-infinite homogeneous mirror, heated and interrogated by a Gaussian beam partially absorbed on the mirror surface, is calculated by Braginsky, Gorodetsky and Vyatchanin~\cite{Brag1} (BGV). A more conclusive form of the photothermal noise spectral density is given by Cerdonio {\it et al.}~\cite{Cerdonio1} in the framework of an extended analysis of thermoelastic noise sources. Their expression can be transformed to give the response to a modulated absorbed power $\delta P_{abs}$, that in the Fourier space reads
\begin{equation}
\delta X(\omega)=-\frac{\alpha(1+\sigma)}{\pi\lambda}\,G\left(\frac{\omega}{\omega_0}\right)\,\delta P_{abs}(\omega)
\label{XBRG}
\end{equation}
($\delta X(\omega)$ is the Fourier transformed of $\bar{u}(t)$) where $\alpha$, $\sigma$, $\lambda$ are respectively the thermal expansion coefficient, the Poisson ratio, and the thermal conductivity of the mirror; $\omega_0=\lambda/\rho C r_0^2$, where $\rho$ is the density and $C$ the specific heat capacity; the shape of the response is
\begin{equation}
G(\Omega) = \frac{1}{\pi} \int_0^\infty \!\!\!\! \ud u \int_{-\infty}^\infty \!\!\! \ud v \;\frac{u^2 e^{-u^2/2}}{(u^2+v^2)(u^2+v^2+i \Omega)} \qquad .
\label{G}
\end{equation}
The requirements for the validity of the above expression (e.g., a short mean free-path of the phonons, and a large sound velocity) are discussed in details in Refs.~\cite{Brag1,Cerdonio1}.

At high frequencies ($\omega \gg \omega_0$, the so-called adiabatic limit), $G(\Omega)$ tends to $1/i\Omega$. This regime occurs when the thermal diffusion length in one modulation period is much smaller than the beam radius. At even higher frequencies, when this length is comparable with the coating thickness, the approximation of homogeneous mirror fails and a more complete model is necessary. 

For an immediate understanding of the physics involved, it is useful to consider a one-dimensional model, starting from the thermal conduction equation that reads
\begin{equation}
\rho C\frac{\partial T}{\partial t}-\lambda\frac{\partial^{2} T}{\partial z^{2}}=Q(t,z)
\end{equation}
where $T$ is the temperature, the mirror occupies the space with $z>0$, and $Q$ is the absorbed power. In the frequency domain, the above equation can be written as
\begin{equation}
\left(k^{2}-\frac{\ud^{2}}{\ud z^{2}}\right)\tilde{T}=\tilde{Q}/\lambda
\label{thdiff}
\end{equation}
where we have introduced a complex wave number
\begin{equation}
k=\sqrt{\frac{i\,\omega\rho C}{\lambda}}  \qquad .
\label{k}
\end{equation}
The inverse of the absolute value of $k$ gives the penetration depth of the thermal wave.

In the case of an absorbed intensity $\tilde{I}_{abs}$ on the surface at $z=0$, the boundary  condition is
\begin{equation}
-\lambda\frac{\ud \tilde{T}}{\ud z}(0)=\tilde{I}_{abs}
\label{bound1}
\end{equation}
and, in a multi-layer medium, at each interface $z_i$ the continuity conditions for temperature and heat flux are
\begin{equation}
\tilde{T}(z)\big|_{z_i^{-}}=\tilde{T}(z)\big|_{z_i^{+}}\qquad\lambda\frac{\ud \tilde{T}}{\ud z}(z)\bigg|_{z_i^{-}}=\lambda\frac{\ud \tilde{T}}{\ud z}(z)\bigg|_{z_i^{+}}  \qquad .
\label{bound2}
\end{equation}

The general solution of the homogeneous equation associated with the (\ref{thdiff}) is
\begin{equation}
\tilde{T}=Ae^{kz}+Be^{-kz}   \, .
\label{exp}
\end{equation}

In the case of a mono-layer with thickness $d$ and associated wave number $k_c$, lying on a substrate with wave number $k_s$, the temperature is given by Eq.~(\ref{exp}) on the layer and by 
\begin{equation}
B_{s} \exp[-k_s (z-d)]
\label{exp1}
\end{equation}
on the substrate, where the coefficients $A$, $B$, and $B_s$ are derived from the boundary conditions (\ref{bound1}) and (\ref{bound2}). The temperature on the layer can also be written in a compact form as
\begin{equation}
\tilde{T}(z)=\frac{\tilde{I}_{abs}}{\lambda_c k_c}\frac{\sinh[k_c (d-z)]+\mathcal{R}\cosh[k_c (d-z)]}{\cosh[k_c d]+\mathcal{R}\sinh[k_c d]}
\label{Tlayer}
\end{equation}
where
\begin{equation}
\mathcal{R}=\frac{k_c\lambda_c}{k_s\lambda_s}=\sqrt{\frac{\lambda_{c}\rho_{c}C_{c}}{\lambda_{s}\rho_{s}C_{s}}}
\label{r}
\end{equation}
and the subscripts $s$ and $c$ distinguish respectively the thermal parameters of the substrate and the layer. The pre-factor determining the temperature in the substrate is
\begin{equation}
B_s=\frac{\tilde{I}_{abs}}{\lambda_s k_s}\frac{1}{\cosh[k_c d]+\mathcal{R}\sinh[k_c d]}  \,\,\, .
\label{Bs}
\end{equation}
We remark that Eq.~(\ref{Bs}) differs from the solution in the absence of the layer (i.e., with $d=0$) by a filter function
\begin{equation}
F= \frac{1}{\cosh[k_c d]+\mathcal{R}\sinh[k_c d]}
\label{Filtro}
\end{equation}
which describes the modification of the heat flux caused by its transmission through the mono-layer.

The filter function can be generally defined as
\begin{equation}
F=-\lambda\frac{\ud\tilde{T}}{\ud z}(d)\cdot\frac{1}{\tilde{I}_{abs}}
\label{Filtrogen}
\end{equation}
where in general the heating power can be absorbed both on the surface and within the medium:
\begin{equation}
\tilde{I}_{abs}=-\lambda\frac{\ud\tilde{T}}{\ud z}(0)+\int_0^d \tilde{Q} \,\ud z \qquad .
\label{Iabsgen}
\end{equation}

As suggested by BGV, the mirror surface deformation can be assumed to be
\begin{equation}
\delta\tilde{u}_z=-\int_0^{\infty}\alpha(z) \tilde{T}(z)\ud z  \, .
\label{uz}
\end{equation}
In this way, the one-dimensional solution differs from the adiabatic approximation only for the lacking term $(1+\sigma)$, if one uses for $\tilde{I}_{abs}$ the peak intensity $\delta P_{abs}/(\pi r_0^2)$. 

It is useful to notice that integrating  Eq.~(\ref{thdiff}) along an homogeneous layer and using Eqs.~(\ref{Filtrogen}) and (\ref{Iabsgen}) we obtain
\begin{equation}
k^2\lambda\int_0^d \tilde{T}=\tilde{I}_{abs}\left(1-F\right) \qquad .
\label{thexpc}
\end{equation}
This expression can be used to write the surface displacement in a compact form, even without performing the direct integration of the temperature along the layer. Indeed, calculating the mirror thermal expansion according to Eq.~(\ref{uz}) and using Eq.~(\ref{thexpc}) and the exponentially decreasing temperature in the substrate, with the pre-factor of Eq.~(\ref{Bs}), we obtain
\begin{equation}
\delta\tilde{u}_z=\left[\frac{\alpha_{c}}{\lambda_{c}k_c^2}\left(F-1\right)- \frac{\alpha_{s}}{\lambda_{s}k_s^2}F\right]\tilde{I}_{abs}     \qquad .
\end{equation}

The above equation is correct in the one-dimensional approximation, that can be relaxed assuming a Gaussian beam with radius $r_0\gg d$, an absorbed intensity $\tilde{I}_{abs}=\delta P_{abs}/(\pi r_0^2)$, and correcting the expansion coefficient of the substrate with the factor $(1+\sigma)$. Observing that $k^2=(i/r_0^2)(\omega/\omega_0)$ and defining $\omega_{c,s}$ similarly to $\omega_0$, we can write
\begin{equation}
\delta X(\omega)=\left[\frac{\alpha_{c}}{\pi\lambda_{c}}\frac{\omega_c}{i \omega}\left(F-1\right)-\alpha_{s}\frac{(1+\sigma_s)}{\pi\lambda_{s}}\frac{\omega_s}{i \omega}F\right]\delta P_{abs}(\omega)   \, .
\label{X1dim}
\end{equation}
On the right side of the above expression, the first term is due to the expansion of the layer, while in the second term we see the expansion of the substrate caused by the filtered thermal flux. Besides the filter, the second term coincides with the adiabatic limit of Eq.~(\ref{XBRG}), as in the case of an homogeneous mirror. A smooth transition to the homogeneous, 3-dimensional model  can therefore be simply obtained by replacing the factor $(\omega_s/i \omega)$ with the function $G(\omega/\omega_s)$ defined in equation (\ref{G}).

A more rigorous expression for $\delta X (\omega)$ is calculated in the Appendix A using a 3-dimensional formalism for both the thermal conduction and the elastic equations. The calculation considers an half-infinite mirror with a superimposed layer, thus neglecting the finite mirror dimensions. The result is summarized in Eqs.~(\ref{B41}-\ref{B46}).

One further thermo-optical phenomenon that must be considered is the thermo-refractive effect: heating changes the optical length of the coating layers, thus modifying the phase of the reflected field and therefore the position of the effective reflection plane with respect to the physical position of the mirror surface. We remark that an increased optical length gives a longer effective cavity (while the thermal expansion brings to a shorter physical cavity), therefore the thermo-refractive effect has in general an opposite sign with respect to photothermal  expansion. The attention to the thermo-refractive effect has been drawn by Braginsky \emph{et al.} \cite{Brag2000}, who identify  it as source of thermo-elastic noise.  A correct expression for the thermo-refractive coefficient in the case of our interest (i.e., a coating starting from the high index layer, without cover layer) is reported in Ref. \cite{Gorodetsky}.
Since the effect is mainly due to the first few layers, we can adopt the simplified expression valid for infinite number of layers and using the temperature at the coating surface:
\begin{equation}
\delta \tilde{u}_z = \lambda_{vac} \beta_{eff} \tilde{T}(0)
\label{uzthr}
\end {equation}
where the thermo-refractive coefficient is
\begin{equation}
\beta_{eff} = \frac{1}{4} \frac{B_H+B_L}{n_H^2-n_L^2}
\label{betaeff}
\end {equation}
with $B_{H,L} = \frac{\partial n_{H,L}}{\partial T}+\alpha_{H,L} n_{H,L}$ \cite{Evans}.
In the framework of a one-dimensional model, the temperature $\tilde{T}(0)$ can be taken from the expression (\ref{Tlayer}) using $z=0$. As for the thermal expansion, the contribution to the signal $\delta X_{tr}(\omega)$ is calculated using the spatial average given by Eq. (\ref{ut}). A more accurate calculation in a three-dimensional framework is reported in Appendix A, and the result is summarized in Eq.~(\ref{B48}).

So far we have only considered surface absorption. In fact, the light is absorbed in the coating, where the electromagnetic field forms a standing wave, with nodes/antinodes in correspondence of the interfaces between layers and a rapidly decreasing amplitude. Moreover, the absorption is in general different in the two materials composing the coating. The exact solution of the field within the coating can be calculated using a characteristic matrix formalism \cite{BW}. For a high-reflectivity coating (with many layers), the average intensity in the first layer is reduced with respect to the intracavity value by a factor very close to the refractive index $n_H$ of the material (in our case, Ta$_2$O$_5$). In each following layer, the intensity is further reduced with respect to the previous one by a factor $n_L/n_H$, where $n_L$ is the index of the second material (for us, SiO$_2$). If we consider that each layer is long $(0.25 \, \lambda_{vac}/n_{H,L})$, where $\lambda_{vac}$ is the vacuum optical wavelength for which the coating is designed, we see that the absorption in the first period is $P_{cav} \cdot 0.25 \, \lambda_{vac}(1/l_H+1/l_L)/n_H^2$, where $l_{H} (l_{L})$ is the extinction length in the H (L) material, and $P_{cav}$ is the intracavity power. In each following period, the absorption is reduced by a factor $(n_L/n_H)^2$, therefore the total absorption in one mirror is $P_{abs}\simeq Abs P_{cav}$ where
\begin{equation}
Abs= 0.25 \, \lambda_{vac}\left(\frac{1}{l_H}+\frac{1}{l_L}\right)\frac{1}{n_H^2-n_L^2}  \qquad .
\end{equation}
For our refractive index values of $n_H = 2.05$ and $n_L = 1.46$, about 74$\%$ of the absorption occurs in the first two periods. In brief, we observe that a) in the two mirrors the absorption is the same, since the kind of substrate and the number of layers are not critical parameters to determine the intensity decay within the coating; b) the two material composing the coating have the same weight in evaluating the total absorption, i.e., the absorption in each material is just determined by its extinction length, independently from  its thickness; c) for our parameters, the approximation of surface absorption is well satisfied since, even at the highest experimental frequencies, the thermal penetration depth corresponds to several periods.

\section{Experiment}

\subsection{Apparatus and procedures}

The optical cavity is made with a OFC (Oxygen Free Copper) spacer, a Corning 7980 fused silica \cite{corning} concave mirror (50~mm radius) as input mirror and a flat, silicon rear mirror. Mirrors diameter and thickness are respectively 12.7~mm and 6~mm. The cavity optical length is 10.7~mm, the beam waist on the flat mirror is 0.083~mm and the corresponding beam size on the input surface is  0.094~mm. The input mirror has a coating composed of alternate quarter-wave layers of Ta$_2$O$_5$ and SiO$_2$ (their physical thickness is respectively 130~nm and 180~nm). The overall number of layers is 29, giving a transmission of about 100~ppm and a total coating thickness of about 4.5~$\mu$m. The rear mirror has a similar coating, with 34 layers (5.3~$\mu$m) and a lower transmission ($\sim 10$ ppm).

The cavity spacer is a thick box, with just two 10~mm diameter holes for the optical access, and few smaller holes for evacuating. It is fixed in thermal contact with the cold finger inside a continuous flow $^{4}$He cryostat which is evacuated down to $5\cdot10^{-4}$~Pa.

Temperature sensors are placed on the cold finger and on the cavity spacer. To assure an efficient cooling and temperature stability of the cavity mirrors, they are mounted inside the OFC box. A spring keep them in contact with the box through their rear, flat surface. The standard thermal shield of the cryostat has four, 25~mm diameter access holes. We have further reduced the apertures to few mm, with aluminum foil. As a consequence,  the mirrors `see' the room temperature through couples of holes (on the spacer and the shield), under a small solid angle.

\begin{figure}[t]
\begin{center}
\includegraphics*[width=1.0\columnwidth]{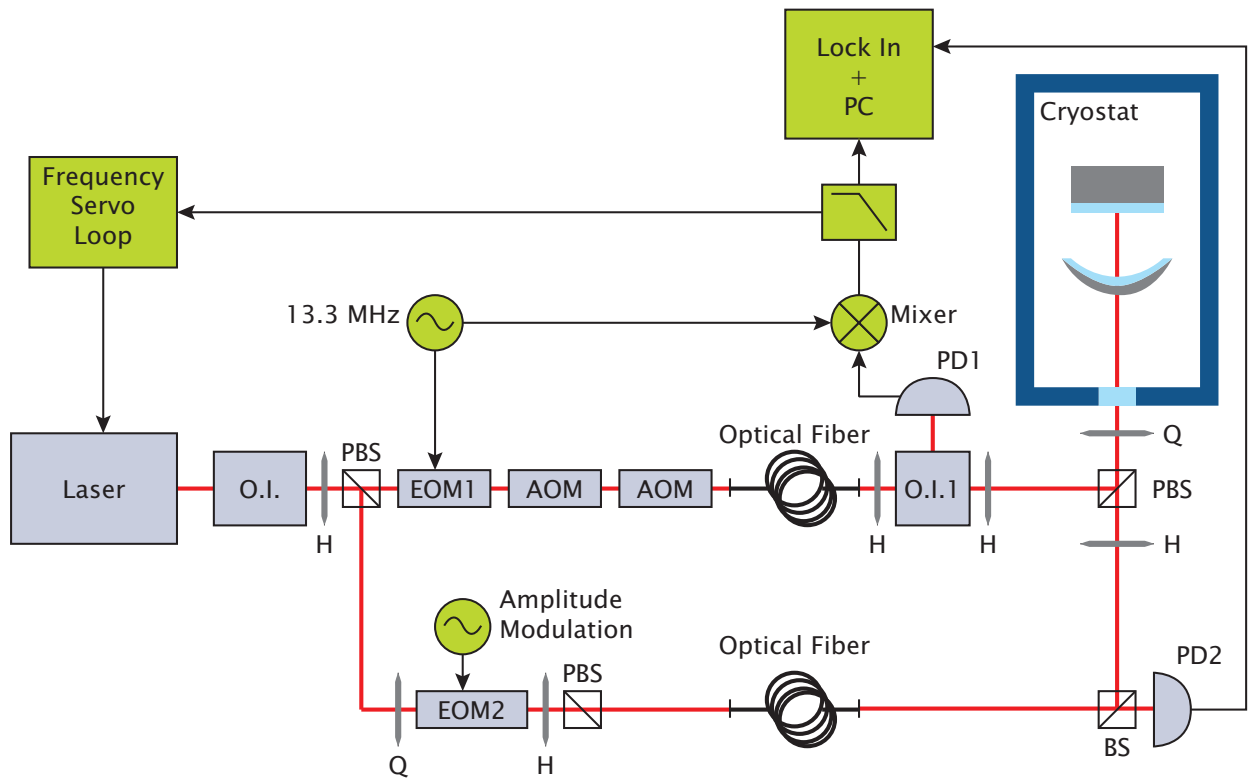}
\end{center}
\caption{Scheme of the experimental apparatus. O.I.: optical isolator; AOM: acousto-optic modulator;
EOM: electro-optic modulator; H: half-wave plate; Q: quarter-wave plate; PD: photodiode; PBS: polarizing beam-splitter; BS: beam-splitter.}
\label{setup}
\end{figure}

The experimental setup for the measurement of the photothermal response is sketched in Fig.~(\ref{setup}). The light source is a cw Nd:YAG laser operating at 1064~nm. After a 40~dB optical isolator, the laser radiation is split into two beams. On the first one (probe beam), a resonant electro-optic modulator (EOM1) provides phase modulation at 13.3~MHz with a depth of about 1~rad used for the Pound-Drever-Hall~\cite{Drever} (PDH) detection scheme. The probe beam is then frequency shifted by means of two acousto-optic modulators (AOM). We use respectively the +1 and -1 diffracted orders of the AOMs, so that the total frequency displacement corresponds to the difference of the AOM frequencies and can be tuned by several MHz around zero.
The intensity of the second beam (pump beam) is controlled by an electro-optic modulator (EOM2) followed by a polarizing beam splitter. Both beams are sent to the second part of the apparatus by means of single-mode, polarization maintaining optical fibers. Part of the pump is then detected by a photodiode (PD2) for monitoring its amplitude modulation, while the probe is sent through a second optical isolator (O.I. 1).
The two beams are overlapped with orthogonal polarizations in a polarizing beam-splitter and sent to the optical cavity. The reflected probe beam, on its back path, is deviated by the input polarizer of the O.I.1 and collected by a photodiode (PD1) for the PDH detection. The pump, slightly detuned with respect to the cavity resonance, is used as excitation beam in the thermo-optical response measurements. Such frequency splitting between
pump and probe allows to eliminate any spurious interference and reduce the cross-talk between them in the photo-detection. A quarter-wave plate before the cryostat allows to compensate cavity birefringence observed at low temperature and probably due to stress on the mirrors generated by differential contractions.
The laser is weakly locked on the cavity using the PDH signal while the pump beam is amplitude modulated at different frequencies. The extraction of the photothermal response is obtained from the same PDH signal, that is acquired by a lock-in amplifier which discriminates the signal component synchronous with the modulation. The depth and phase of the intracavity power modulation is calculated from the signal of PD2, which is also acquired and de-modulated by the lock-in amplifier, by taking into account the mode matching and the cavity coupling factor. The latter is inferred from the depth of the dip in the reflected beam when scanning on the resonance, and it is consistent with the independent measurements of cavity Finesse and input mirror transmission.

The PDH signal is calibrated in a separate run, stopping the pump beam and modulating the frequency of the laser through its internal piezoelectric crystal, with a depth smaller than the Fabry-Perot cavity linewidth. The usual lock-in acquisition and phase-sensitive detection is applied to the PDH signal. The frequency modulation rate had been previously measured by observing the sidebands at 13.3~MHz with a second, high-Finesse optical cavity. The calibration procedure, performed varying the modulation frequency, allows also to correct the data for the servo loop. The locking bandwidth is of the order of 1~kHz.

Typical values of the optical parameters are: a mode-matching of 80$\%$; a Finesse of 25000 (linewidth $\sim 475$~kHz); an intracavity optical power $P_{cav}$ equal to 7000 times the input pump power $P_{in} \sim 2$~mW. The input probe beam power is approximately 0.2~mW.

\subsection{Experimental results and data analysis}
The photothermal response to intensity modulation of the intracavity field has been investigated both at room and cryogenic temperatures in the 10~Hz~$\div$~100~kHz frequency range. The measured signal is reported in terms of changes of the optical path divided by the amplitude of the modulation in the intracavity power, giving a response in m/W.
Previous experiments \cite{DeRosa02,DeRosa06} have shown that, at low frequencies (below $\sim$100~Hz), the photothermal expansion is determined by the substrate contribution. In agreement with the theory of homogeneous mirrors, the curve can be roughly divided into two regions, separated by the characteristic angular frequency $\omega_s$ which depends on the thermal properties of the mirrors material. Well above $\omega_s$, the surface displacement is expected to decrease as the inverse of the frequency while, well below $\omega_s$, it tends to a constant value which determines the absorption coefficient. On the other hand, at higher frequencies (above $\sim$100~Hz), the coating contribution to the total thermal expansion, in particular through the filter function, becomes relevant and the observed behavior differs substantially from the substrate $1/f$ frequency dependence. At even higher frequencies, also the thermo-refractive effect becomes important (in the following, we will use the term `photothermal response' for the overall effect, including the thermo-refractive component). An accurate data analysis through Eq.~(\ref{Xfinal}) and the related expressions in Appendix A allows us to to measure the effective thermal properties of the Ta$_2$O$_5$/SiO$_2$ dielectric layers. Moreover, assuming that the thermal parameters of silica films are not significantly different from those of bulk silica, we will estimate in the next section the properties of Ta$_2$O$_5$ thin film. On this material, to our knowledge, just few, room temperature measurements have been performed.

\begin{figure}[t]
\begin{center}
\includegraphics*[width=0.8\columnwidth]{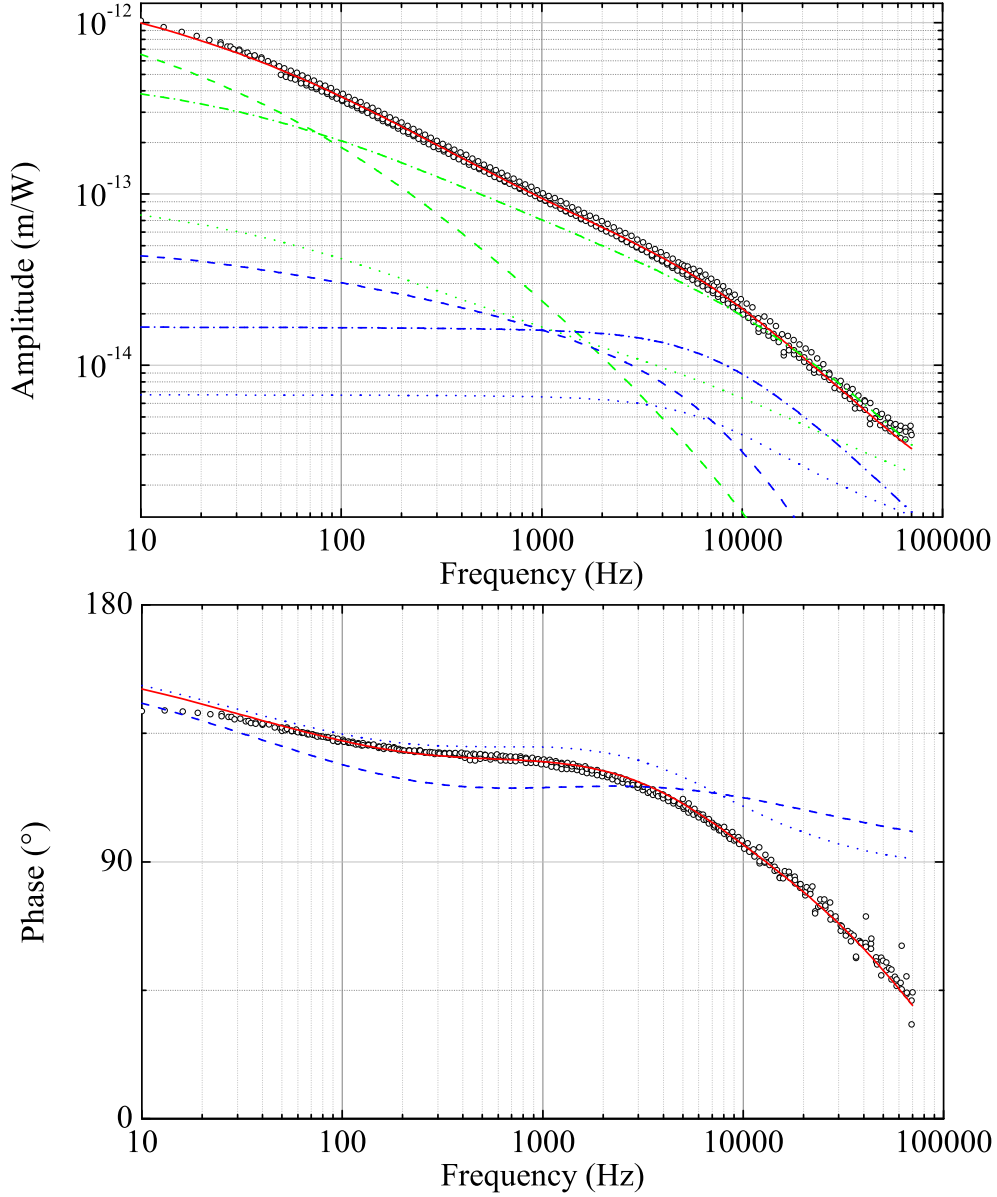}
\end{center}
\caption{a) Amplitude of the photothermal response as a function of
the modulation frequency at room temperature. The solid red line is the fit through the complex response function (\ref{Xfinal}) for both silica and silicon mirrors, keeping fixed the substrate parameters (see Table~I). Other curves refer to the fused silica (green) and silicon (blue) mirrors. Dashed lines: substrate contributions; dash-dotted lines: coating expansion contributions; dotted lines: thermo-refractive contributions.
b) Phase of the photothermal response and its corresponding fit. Blue curves are traced neglecting some effects, in order to show their relevance: the dashed  line corresponds to a response of just homogeneous substrates (calculated from Eq.~(\ref{XBRG}), the dotted line considers just $\delta X_s$ and $\delta X_c$ in Eq.~(\ref{Xfinal}), thus neglecting the thermo-refractive effect.}
\label{fig2}
\end{figure}

The amplitude and phase of the photothermal response at room temperature are shown in Fig.~(\ref{fig2}). The experimental data (amplitude and phase) are globally fitted through the complex response function (\ref{Xfinal}) considering both the silica and silicon substrate contributions and the coating. The fitting function is shown in Fig.~(\ref{fig2}) (red traces), and  Fig.~(\ref{fig2}a) also shows the amplitudes of some specific contributions. In the fitting procedure, the substrates thermal properties (thermal conductivity, specific heat capacity and expansion coefficient) are fixed to values taken from the literature and reported in Table~I. The coating parameters derived from the fit are reported in Table~II (first column). 

The mechanical unbalance parameters $\gamma_1$ and $\gamma_2$ (see Eq.~(\ref{B39a})) are both close to unit and depend on the Young modulus ($E$) and the Poisson factor ($\sigma$) of substrate and coating. We have taken the values of $E$ and $\sigma$ from the literature (they are reported in Tables I and II), including the experimental uncertainty for the coating, and a spread due to the dependence on the direction with respect to the crystal axes for Si, since we have no information on their orientation in the mirror. The calculated parameters are $\gamma_1 = 1.36\pm0.05$ and $\gamma_2 = 1.53\pm0.12$ for the silica mirror; $\gamma_1 = 0.97\div0.99\pm0.02$ and $\gamma_2 = 0.58\div0.80\pm0.06$ for the silicon mirror. The uncertainty in the mechanical parameters gives a small spread in the fitted parameters (below the statistical error given by the fit), that we have included in their quoted uncertainty in Table II. 

The absorption coefficients of the two mirrors are kept as different parameters even if the two coatings are very similar (their slightly different thickness is not expected to have a great influence on the absorption), since they have been provided in different batches and we have no precise information on the post-coating annealing procedure that can be different for the two mirrors.

The coating parameters $(\rho_c C_c)$, $\lambda_c$, $\alpha_c$, and $\beta_{eff}$ are determined by the fit. 
Its results are reported in Table II together with their uncertainties derived from the 68~\% confidence intervals, slightly increased to account for the spread in the used values of Young modulus and Poisson factor. 

At frequencies below 100~Hz the silica substrate contribution (whose characteristic frequency is $\omega_s^{FS} \sim 2\pi \cdot 27$~Hz) is dominant while at higher frequencies the photothermal response is dominated by the silicon substrate ($\omega_s^{Si} \sim 2\pi \cdot 4200$ Hz) and the coating. 
The importance of the coating is not restricted to its direct contribution to the photothermal expansion (dash-dotted curves), but it is evident above all in the filter function that strongly affects the total response. This is put into evidence in Fig. (\ref{fig2}b), where the phase data are also compared with a curve obtained form Eq.~(\ref{XBRG}), thus neglecting the filter function contribution (dashed curve). In this case, the disagreement between theory and experiment is relevant even at low frequencies and would lead to unreliable estimations of both the substrate and coating parameters. The phase of the response is sensitive  also to the thermo-refractive effect, which seems negligible if one just consider its amplitude. This is underlined again in Fig. (\ref{fig2}b), where the dotted curve neglects $\delta X_{tr}$ in Eq.~(\ref{Xfinal}). By considering the complete model (red curve), instead, the agreement is excellent in the whole frequency range.

We observe that the coating on silicon mirror displays a lower absorption. The absorption coefficients that we have obtained are a bit higher than previously reported~\cite{DeRosa06}, yet remaining within few ppm. We remark however than the quoted uncertainty is reliable for comparing the two mirror, but the estimate of the overall absorption is conditioned by the evaluation of the intracavity power that is hardly very accurate.

\begin{figure}[t]
\begin{center}
\includegraphics*[width=0.8\columnwidth]{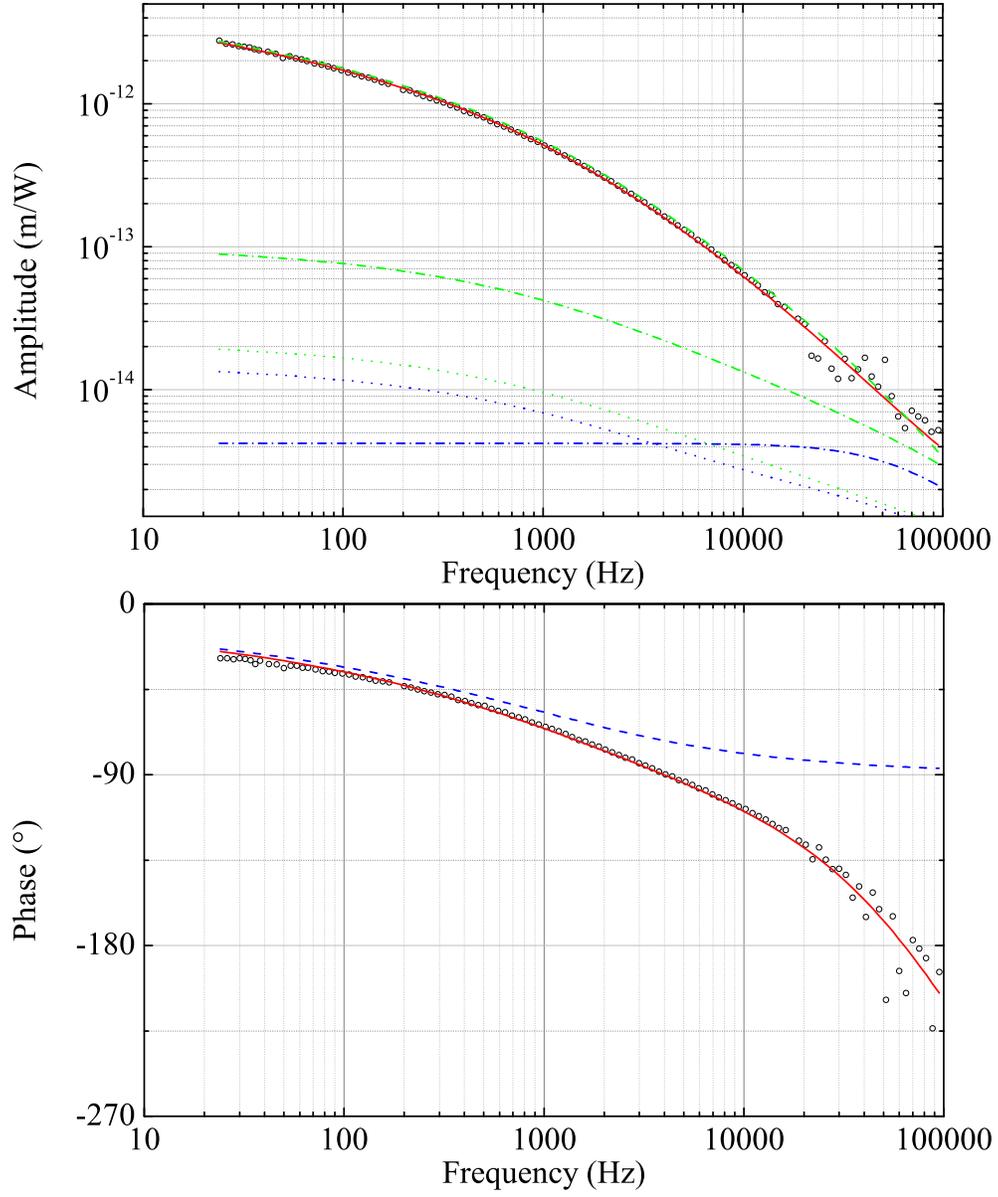}
\end{center}
\caption{The same as in Fig. (\ref{fig2}), at cryogenic temperature.}
\label{fig3}
\end{figure}

The measurements have been repeated at low temperature, with the cavity spacer at $4.5$~K. The results are shown in Fig.~(\ref{fig3}), together with the fitting functions, and the derived parameters are reported in Table II (second column). Around this temperature, the photothermal response is dominated by the fused silica substrate at all frequencies. 

We remark that, at low frequencies, the phase curves reported in Fig.~(\ref{fig2}b) and Fig.~(\ref{fig3}b)
clearly show that the photothermal response at room and cryogenic temperature are $180^{o}$ out of phase. This is due to the sign change in the fused silica expansion coefficient, which becomes negative below $\sim 170$~K. The effect of the coating filter function on the photothermal response phase is even more important than at room temperature, as evidenced in Fig.~(\ref{fig3}b).

For what concerns the silicon substrate, on one hand the silicon thermal diffusion length becomes larger than the mirror size for frequencies below few kHz, thus invalidating our theoretical model (the temperature of the silicon substrate could be considered as homogeneous, instantaneously determined by the heat absorption and the conduction through the copper spacer); on the other hand, its thermal expansion coefficient becomes so low that it can be completely neglected. Some figures justifying this assumption are reported in Table~I.

Concerning the thermal parameters of the silica substrate, we remark that expansion coefficient and capacitance have a strong dependence on temperature. Since the real temperature of the mirrors is also determined by their thermal contact with the copper spacer, we have kept it as free parameters in the fit, using the known temperature dependence of $C_{SiO_2}$ and $\alpha_{SiO_2}$ (reported in the caption of Table~I). The derived temperature of the mirrors is indeed $10.9\pm0.3$~K, giving for bulk silica  $\alpha_{SiO_2} = -0.267\pm0.012\cdot10^{-6}$~K$^{-1}$ and $C_{SiO_2} = 5.2\pm0.5$~J/kg~K.

The mechanical parameters ($\gamma_1$, $\gamma_2$, $\sigma$) are kept at their room temperature values (their weak temperature dependence is negligible in our work). We have verified that their uncertainty has negligible effects on all the fitted parameters. 

The contribution of the coating on silicon is taken into account, imposing 
$\mathcal{R} = 0$ in the relevant equations as justified by the silicon 
thermal parameters reported in Table~I. The ratio between the absorption 
coefficients of the two mirrors is fixed at the value found at room 
temperature (i.e., $Abs_{Si}=0.44\,Abs_{SiO2}$). The $\sim50\%$ uncertainty 
in this ratio adds a small error to some fitted parameters (well below the 
corresponding statistical errors form the fit), that is included in the 
uncertainties quoted in Table~II. The absorption in the silica mirror results to be lower than at room temperature. However, we remark again that the quoted uncertainty in the absorption coefficients only reflects the statistical error, while absolute intracavity power calibrations (giving additional systematic uncertainty) are different between runs at room and cryogenic temperature due to the necessary re-alignment. 

The thermo-refractive effect at cryogenic temperature is too small to allow a reliable independent estimate of both $\alpha_c$ and $\beta_{eff}$ from our data. Yet, it cannot be completely neglected. Therefore, we have left $\alpha_c$ as free parameter and used the expression $\beta_{eff}$ of Eq.~(\ref{betaeff}) as follows: for the $B_L$ parameter (concerning silica) we have used the silica bulk values of $\alpha$ and $\partial n/\partial T$ reported in Table~I; the expansion coefficient of tantala is derived from Eq.~(\ref{avexp}) using the free parameter $\alpha_c$ and the bulk expansion coefficient of silica; we have varied the thermo-optic coefficient $\partial n/\partial T$ of tantala between 0 and twice the silica value. The spread in this coefficient has negligible effect in all the fitted parameters except for $\alpha_c$, where the spread is comparable to the statistical uncertainty of the fit. Since our assumption on the range of the tantala thermo-optic coefficient is rather arbitrary, we have quoted separately its effect on $\alpha_c$ in Table~II to allow a reliable evaluation of our results once an independent measurement of $\partial n/\partial T$ for tantala is provided.

\subsection{Discussion}
\label{Discussion}
As discussed before, the fitting procedure allows the estimate of the coating effective thermal conductivity $\lambda_c$, the product $\rho_c C_c$, the averaged thermal expansion coefficient $\alpha_c$ and the thermo-refractive coefficient $\beta_{eff}$. The values of these effective parameters are the main result of our work, and they can be directly used in most evaluations of noise sources and, in general, physical phenomena related to the high reflectivity coating in interferometers. In this section, we further discuss our results with the aim of extracting the parameters of the materials composing the coating, and comparing with existing literature.

The models used in our data analysis consider indeed an homogeneous coating layer. In a complete description of the film behavior, the real parameters of the layers would enter with several different combinations that should be individually averaged over the coating. However, to extract some estimates of the layers parameters, we adopt the simplified description suggested, e.g., in Ref. \cite{Fejer}, that takes volume averages such as
\begin{equation}
\frac{1}{\lambda_c}=\frac{1}{d_p}\left(\frac{d_H}{\lambda_H}+\frac{d_L}{\lambda_L}\right)
\label{parametrikeff1}
\end{equation}
\begin{equation}
\rho \,C_c  = \frac{1}{d_p}\left(d_H \rho_H \,C_H + d_L \rho_L\, C_L\right)
\label{parametrikeff1a}
\end{equation}
\begin{equation}
\label{avexp}
\alpha_c = \frac{\alpha_H  d_H+\alpha_L  d_L}{d_p} \qquad .
\end{equation}
\begin{equation}
d_p=d_H+d_L
\label{parametrikeff2}
\end{equation}
where the subscripts `$H$' and `$L$' refer to the two consecutive layers composing one period of the coating. Eqs.~(\ref{parametrikeff1}-\ref{parametrikeff2}) and (\ref{betaeff}) allow us to extract the parameters of the materials composing the film from our experimental results. 

We start our discussion from the room temperature data. Ref. \cite{Grilli} presents a review of conductivity measurements in SiO$_2$ and  Ta$_2$O$_5$ films, for different thickness and deposition techniques. The range of results is very wide, extending between 0.017 and 0.73~W/mK for silica, and between 0.026 and 15~W/mK for Ta$_2$O$_5$. If we assume, in spite of that, for the SiO$_2$ layers the bulk parameter values reported in Table~I, from Eq.~(\ref{parametrikeff1}) we obtain for Ta$_2$O$_5$ a thermal conductivity of $\lambda_{Ta_2 O_5}$=0.4~W/mK. We stress that such estimate should be taken \emph{cum grano salis}, since conductivity in thin films can strongly depend on the film structure; the most meaningful parameter remains $\lambda_c$. 

Concerning the heat capacity, it can be estimated using the bulk parameters for silica, the measured density of 7200~kg/m$^3$ for the Ta$_2$O$_5$ film \cite{Flaminio}, and the bulk tantala capacity of 306~J/kg~K~\cite{Kelley}. We calculate from Eq.~(\ref{parametrikeff1a}) a value of $(\rho_c C_c)\simeq1.9\cdot10^6$~J/m$^3$~K. The result of our fit is $(\rho_c C_c)= 1.4 \pm 0.3\cdot10^6$~J/m$^3$~K, that is compatible (within two standard errors), while a bit lower, with the previous estimate. 

The thermal expansion coefficient of Ta$_2$O$_5$ can be extracted through Eq.~(\ref{avexp}), using again the bulk data for silica. We obtain $\alpha_{Ta_2 O_5}$=4.8$\pm0.7\cdot 10^{-6}$ $\rm{K^{-1}}$ to be compared with the values of -4.4~$10^{-5}$ $\rm{K^{-1}}$ \cite{inci}, 2.42~$10^{-6}$ $\rm{K^{-1}}$ \cite{tien} and 5~$10^{-6}$ $\rm{K^{-1}}$ \cite{bg} measured in previous experiments. Ref.~\cite{tien} also report a value of 3.6~$10^{-6}$ $\rm{K^{-1}}$ obtained from the analysis of data in the literature. Our value is closer to those reported in Refs.~\cite{tien,bg} than to the one obtained by Inci~\cite{inci} which, apart the unexpected negative sign, is more than a factor of 10 larger in absolute value. 
As discussed in Ref.~\cite{bg}, the variability between these independent measurements could be due to the fact that the properties of tantalum pentoxide layers might depend on the procedure of the layer deposition. 

An accurate measurements of the refractive index of bulk silica at temperature varying from 30 to 310~K is reported in Ref.~\cite{Leviton}. From its data, we infer (at $\lambda_{vac}=1064$nm) $\partial n/\partial T=8.57\cdot 10^{-6}$~K$^{-1}$ at 300~K, and $\partial n/\partial T \simeq(8+3.3 T)\cdot 10^{-8}$~K$^{-1}$ at low temperature. The room temperature value is compatible with $\partial n/\partial T=1.2\cdot 10^{-5}$~K$^{-1}$ measured at 632.8~nm for a sputtered SiO$_x$ film in Ref.~\cite{chu}. Using Eq.~(\ref{betaeff}) and the bulk data for silica, we obtain for the tantalum pentoxide $\partial n/\partial T + n \alpha =27\pm 5\cdot 10^{-6}$~K$^{-1}$, and from our value of $\alpha_{Ta_2 O_5}$ we calculate ($\partial n/\partial T)_{Ta_2 O_5} = 17\pm 7 \cdot 10^{-6}$~K$^{-1}$. Previous measurements of ($\partial n/\partial T)_{Ta_2 O_5}$, performed at 632.8~nm by a group in Taiwan, gives 2.3$\cdot 10^{-6}$~K$^{-1}$~\cite{chu} and, later, 3.64$\cdot 10^{-6}$~K$^{-1}$ and 7.89$\cdot 10^{-6}$~K$^{-1}$~\cite{Cheng1999} (depending on the thermal stress). A further value of 
14$\cdot 10^{-6}$~K$^{-1}$ is mentioned in \cite{Evans} as a private communication. 

At low temperature, the literature on the coating properties is very poor. By using, as before, the expressions (\ref{parametrikeff1})-(\ref{parametrikeff2}) together with the bulk properties of Table~I for silica, we deduce the following data for the film of tantalum pentoxide: the expansion coefficient is $\alpha_{Ta_2 O_5}$=0.58$\pm0.04\pm0.05\cdot 10^{-6}$ $\rm{K^{-1}}$ (the first quoted uncertainty is due to the spread in the assumed thermo-optic parameter of tantala between 0 and 2~$\times$~$\partial n/\partial T_{SiO_2}$, with $\partial n/\partial T_{SiO_2} = 0.44\cdot10^{-6}$ at 10.9~K; the second quoted uncertainty is the statistical error from the fit); the thermal capacitance is $(\rho\,C)_{Ta_2 O_5} = 2.2\pm0.7\cdot10^{4}$~J/K~m$^3$ and, using a density of 7200~kg/m$^3$, we calculate $C_{Ta_2 O_5} = 3.1\pm1.0$~J/kg~K; the thermal conductivity is $\lambda_{Ta_2 O_5}=5\pm3\cdot10^{-2}$~W/mK.

\section{Conclusions}

We have measured the variation of the effective reflecting surface in a high reflectivity mirror, caused by a modulation in the residual absorbed power. Such complex response is due to photothermal and thermo-refractive effects in the mirror substrate and coating, and our measurements allow a characterization of these effects both at room and at cryogenic temperature. An accurate knowledge of such phenomena is particularly relevant in experiments aiming to the highest sensitivity in displacement, position and force measurements, for instance in the fields of gravitational wave detection and in quantum optics. Most of the planned future experiments of this kind will address cryogenics to reduce Brownian noise, and part of the present work is indeed the focused on cryogenic temperature, where measurements of the kind here presented were still missing.

Our experimental results are well described using a theoretical model that we have expressly developed, and the comparison allows us to extract meaningful effective thermo-mechanical parameters of the high reflectivity coating. This is the most important aspect of our work, since a) the overall coating behavior is not obviously deduced from the properties of the composing materials (in particular concerning thermal expansion and conductivity); b) the parameters that we have measured can be used to calculate weak but important noise contributions such as thermodynamic fluctuations or, of course, photothermal and thermo-refractive noise. 

We underline that, while specific experiments on the different oxides in the bulk or thin-film form can give more accurate results concerning the material properties, our setup has the  added value of characterizing the mirrors (and coating) as they are used in high sensitivity experiments, thus describing their real overall behavior. Moreover, from the effective coating parameters we could still estimate some properties of the dielectric films composing the coating. In particular, literature on previous measurements were rather poor for tantalum pentoxide at room temperature, and, in our knowledge, completely lacking at cryogenic temperature.

\section{Acknowledgments}

This work was partially funded by the European Union ILIAS Project (No. RII3-CT-2003-506222).

\section{Appendix A}

In this Appendix we calculate the temperature profile and the elastic deformation of an half-infinite mirror extended in the half-space $z>0$, composed of a layer of thickness $d$ an an homogeneous substrate, with surface absorption of a Gaussian beam. We will consider an isotropic substrate, neglecting the directional variation of the Si properties. The calculation is performed using Fourier transforms for the temporal and transverse spatial variables, while keeping the dependence on the longitudinal variable $z$. 

\subsection{Thermal equations}

We start by the study of the temperature distribution. We define 
\begin{equation}
\tilde{T}(\omega,\mathbf{r})=\int_{-\infty}^{\infty} \ud t \, e^{-i\omega t}\,T(t,\mathbf{r}) 
\label{B1}
\end{equation}
and
\begin{equation}
\Theta(\omega,\mathbf{k_{\perp}},z)=\int\int_{-\infty}^{\infty} \ud^2 \mathbf{r_{\perp}} \, e^{-i\mathbf{k_{\perp}}\cdot \mathbf{r_{\perp}}}\,\tilde{T}(\omega,\mathbf{r}) 
\label{B2}
\end{equation}
where $\textbf{r}=(x,y,z)$, $\textbf{r}_{\perp}=(x,y)$, and $\textbf{k}_{\perp}=(k_x,k_y)$.
The  equation for the thermal propagation 
\begin{equation}
\frac{\rho C}{\lambda}\frac{\partial T}{\partial t}-\nabla^2 \, T=0 
\label{B3}
\end{equation}
can be written in the time-transformed Fourier space as
\begin{equation}
k^2 \tilde{T}-\nabla^2 \, \tilde{T}=0 
\label{B3a}
\end{equation}
where $k$ is defined in Eq.~(\ref{k}), and in the space- and time-transformed Fourier space as
\begin{equation}
\left(k^2+k_{\perp}^2\right)\Theta-\frac{\partial^{2} \Theta}{\partial z^{2}}=0 \qquad.
\label{B4}
\end{equation}

The absorbed intensity, in the time-transformed Fourier space, is written as
\begin{equation}
\tilde{I}(\omega,\mathbf{r}_{\perp})=\frac{e^{-r_{\perp}^2/r_0^2}}{\pi r_0^2}\,
\delta P_{abs}(\omega)
\label{B5}
\end{equation}
and, in the complete Fourier description where $\mathcal{I}$ is the spatial Fourier transformed of $\tilde{I}$, it becomes
\begin{equation}
\mathcal{I}(\omega,\mathbf{k}_{\perp})=\exp \left(-\frac{k_{\perp}^2 \,r_0^2}{4}\right)\,\delta P_{abs}(\omega) \qquad.
\label{B6}
\end{equation}

The three-dimensional model is now formally equivalent to the one-dimensional model described in Section~\ref{model}. 
The equations (\ref{B4}) is equal to the homogeneous equation associated to Eq.~(\ref{thdiff}), and the boundary conditions (\ref{bound1}) and (\ref{bound2}) are still valid, with the following list of substitutions:
\begin{equation} 
\tilde{T} \to \Theta
\label{B7}
\end{equation}
\begin{equation} 
\tilde{I}_{abs} \to \mathcal{I}
\label{B8}
\end{equation}
\begin{equation} 
k^2 \to i \frac{\rho \, C}{\lambda}\omega+k_{\perp}^2 = k^2 +k_{\perp}^2\, .
\label{B9}
\end{equation}
The discussion on the temperature distribution in the one-dimensional model remains valid too. We can therefore use Eqs.~(\ref{exp}-\ref{Filtro}), with the replacements (\ref{B7}-\ref{B9}) and where $k_s$ and $k_c$ contain the respective thermal parameters. For the sake of clarity, we repeat here explicitly the solutions, in the layer (from Eq.~(\ref{Tlayer})): 
\begin{equation}
\Theta = \frac{\exp \left(-\frac{k_{\perp}^2 \,r_0^2}{4}\right)}{\lambda_c \sqrt{k_c^2+k_{\perp}^2}}\,\frac{\sinh\left[\sqrt{k_c^2+k_{\perp}^2} \,(d-z)\right]+\mathcal{R}\cosh\left[\sqrt{k_c^2+k_{\perp}^2}\, (d-z)\right]}{\cosh\left[\sqrt{k_c^2+k_{\perp}^2}\, d\right]+\mathcal{R}\sinh\left[\sqrt{k_c^2+k_{\perp}^2}\, d\right]}\,\delta P_{abs}(\omega)
\label{B9a}
\end{equation}
and in the substrate (from Eqs.~(\ref{exp1},\ref{Bs})):
\begin{equation}
\Theta = \frac{\exp \left(\frac{k_{\perp}^2 \,r_0^2}{4}\right)}{\lambda_s \sqrt{k_s^2+k_{\perp}^2}}\,\frac{\exp\left[-\sqrt{k_s^2+k_{\perp}^2}\, (d-z)\right]}{\cosh\left[\sqrt{k_c^2+k_{\perp}^2}\, d\right]+\mathcal{R}\sinh\left[\sqrt{k_c^2+k_{\perp}^2}\, d\right]}\,\delta P_{abs}(\omega)
\label{B9b}
\end{equation}
where the reflection coefficient is now
\begin{equation}
\mathcal{R} = \frac{\lambda_c\,\sqrt{k_c^2+k_{\perp}^2}}{\lambda_s\,\sqrt{k_s^2+k_{\perp}^2}} \qquad .
\label{B9c}
\end{equation}

\subsection{Elastic equations}

We now consider the problem of the mirror thermal expansion. The elastic deformation $\textbf{u}$ is ruled by the static elastic equation~\cite{Brag1}
\begin{equation}
\frac{1-\sigma}{1+\sigma}\,\nabla\,(\nabla \cdot \mathbf{u})\,-\,\frac{1-2\sigma}{2(1+\sigma)}\,\nabla\,\wedge (\nabla \wedge \mathbf{u})\,=\,\alpha \, \nabla\,T   \qquad .
\label{B10}
\end{equation} 
The relevant components of the stress tensor are defined as
\begin{equation}
\sigma_{zz} = \frac{E}{1-2\sigma}\left[\frac{\sigma}{1+\sigma}\nabla\cdot \mathbf{u}+\frac{1-2\sigma}{1+\sigma}\frac{\partial u_z}{\partial z}-\alpha T\right]
\label{B10a}
\end{equation} 
\begin{equation}
\sigma_{zx} = \frac{E}{2(1+\sigma)} \left(\frac{\partial u_z}{\partial x}+\frac{\partial u_x}{\partial z}\right)
\label{B10b}
\end{equation}
\begin{equation}
\sigma_{zy} = \frac{E}{2(1+\sigma)} \left(\frac{\partial u_z}{\partial y}+\frac{\partial u_y}{\partial z}\right) 
\label{B10c}
\end{equation}
where $E$ is the Young modulus and $\sigma$ is the Poisson ratio. 

Following BGV, we search a particular solution of Eq.~(\ref{B10}) in the form of a gradient of a potential $\phi$. Inserting $\nabla \phi$ in Eq.~(\ref{B10}), we see that the potential $\phi$ obeys the Poisson's equation
\begin{equation}
\nabla^2 \phi = \frac{1+\sigma}{1-\sigma}\, \alpha \, T  \qquad .
\label{B14}
\end{equation}

We then search a complete solution of the kind
\begin{equation}
\mathbf{u} = \mathbf{v}+\nabla \phi   
\label{B15}
\end{equation}
that is inserted  in Eq.~(\ref{B10}) to show that $\textbf{v}$ satisfies the associated homogeneous equation, that we recast in the form
\begin{equation}
\nabla (\nabla \cdot \mathbf{v})+(1-2\sigma)\nabla^2 \mathbf{v} =0  \qquad .
\label{B16}
\end{equation}

As in the study of the thermal diffusion, we define the time- and transverse spatial Fourier transformed variables $\Phi(\omega,\textbf{k}_{\perp},z)$ of $\phi$; $U_i(\omega,\textbf{k}_{\perp},z)$ and $V_i(\omega,\textbf{k}_{\perp},z)$ of the Cartesian components of $\textbf{u}$ and $\textbf{v}$, respectively, with $i=x,y,z$. The equation (\ref{B16}), in its Cartesian components, can be written in the Fourier space, giving
\begin{equation}
\frac{\partial}{\partial \, z}\left(i k_x V_x+i k_y V_y+\frac{\partial V_z}{\partial \, z}\right)+(1-2\,\sigma)\left(-k_{\perp}^2+\frac{\partial^2 }{\partial \, z^2}\right)V_z=0
\label{B18}
\end{equation}
from the $z-$ component, 
\begin{equation}
i k_x \left( i k_x  V_x +i k_y V_y +\frac{\partial V_z}{\partial \, z}\right)+(1-2\,\sigma)\left(-k_{\perp}^2+\frac{\partial^2 }{\partial \, z^2}\right)V_x=0
\label{B19}
\end{equation}
from the $x-$ component, and the equivalent of Eq.~(\ref{B19}) with exchange of $x$ and $y$ for the $y-$ component.

The axial symmetry of the problem implies that $V_x/k_x=V_y/k_y$. We can therefore define
\begin{equation}
V_r = i\frac{k_{\perp}}{k_x} V_x = i\frac{k_{\perp}}{k_y} V_y
\label{B20}
\end{equation}
that, replaced in Eq.~(\ref{B18}), gives
\begin{equation}
k_{\perp}\frac{\partial V_r}{\partial \, z}+2(1-\,\sigma)\left(-k_{\perp}^2+\frac{\partial^2 }{\partial \, z^2}\right)V_z=0
\label{B21}
\end{equation}
and replaced in either Eq.~(\ref{B19}) or its equivalent with $x\leftrightarrow y$ (the two equations transforms into the same expression) gives
\begin{equation}
-k_{\perp}\frac{\partial V_z}{\partial \, z}+2(1-\,\sigma)\left(-k_{\perp}^2+\frac{\partial^2 }{\partial \, z^2}\right)V_r=0   \qquad .
\label{B22}
\end{equation}
We remark that the condition $V_x/k_x=V_y/k_y$ simplifies very much the calculation. However, one can also avoid to impose it at the beginning, and verify later that it results as a consequence of the boundary conditions.

Eqs.~(\ref{B21}) and (\ref{B22}) form a system of homogeneous, second order differential equations with constant coefficients, in the variable $z$, for the vector of functions $\{V_z,V_r\}$. The solution is a linear combinations of the four vectors $\{e^{k_{\perp} z},-e^{k_{\perp} z}\}$; $\{e^{-k_{\perp} z},e^{-k_{\perp} z}\}$; $\{k_{\perp}\, z \, e^{k_{\perp} z},(4\sigma-3-k_{\perp}\, z )\,e^{k_{\perp} z}\}$; $\{k_{\perp}\, z\, e^{-k_{\perp} z},(4\sigma-3+k_{\perp} \,z )	,e^{-k_{\perp} z}\}$. It is useful to express the solution in different forms for the layer (in $0 < z < d$) and in the substrate (for $z>d$), where we skip the diverging solutions with positive real part of the exponent. Therefore, we write the solutions as
\begin{equation}
V_z = V_c\,\cosh (k_{\perp} z)+V_s\,\sinh (k_{\perp} z)+V_{cc}\,k_{\perp}\,z\,\cosh (k_{\perp} z)+V_{ss}\,k_{\perp}\,z\,\sinh (k_{\perp} z)  
\label{B23}
\end{equation}
\begin{eqnarray}
V_r = -\,V_c\,\sinh (k_{\perp} z)\,-\,V_s\,\cosh (k_{\perp} z)+V_{cc}\,\big[(4\sigma_c-3)\,\cosh (k_{\perp} z)\,-\,k_{\perp}\,z\,\sinh (k_{\perp} z)\big]  \nonumber\\
+V_{ss}\,\big[(4\sigma_c-3)\,\sinh (k_{\perp} z)\,-\,k_{\perp}\,z\,\cosh (k_{\perp} z)\big]
\label{B24}
\end{eqnarray}
in the layer and
\begin{equation}
V_z = V_a\,e^{-\,k_{\perp} (z-d)}+V_b\,k_{\perp}\, (z-d)\,e^{-\,k_{\perp} (z-d)}
\label{B25}
\end{equation}
\begin{equation}
V_r = V_a\,e^{-\,k_{\perp} (z-d)}+V_b\,\big[(4\sigma_s-3)+k_{\perp}\, (z-d)\big]\,e^{-\,k_{\perp} (z-d)}
\label{B26}
\end{equation}
in the substrate. Here $\sigma_c$ and $\sigma_s$ are the Poisson factors respectively in the layer and in the substrate, and $V_c, V_s, V_{cc}, V_{ss}, V_a, V_b$ are functions of $\omega$ and $k_{\perp}$.

For what concerns the particular solution $\nabla \phi$, instead of solving Eq.~(\ref{B14}), we can compare it with  Eq.~(\ref{B3a}) and directly write (in the complete Fourier space)
\begin{equation}
\Phi = \alpha \, \frac{1+\sigma}{1-\sigma}\,\frac{1}{k^2}\, \Theta  \qquad .
\label{B27}
\end{equation}

The definitions of the stress tensor given in Eqs.~(\ref{B10a}-\ref{B10c}) can also be written in the complete Fourier space. Naming $\tilde{\sigma}_{ij}$ the space- and time-Fourier transformed of $\sigma_{ij}$, using Eqs.~(\ref{B15}) and (\ref{B27}), we obtain
\begin{equation}
\tilde{\sigma}_{zz} = \frac{E}{1-2\sigma}\left[k_{\perp}\,\frac{\sigma}{1+\sigma}\,V_r+\frac{1-\sigma}{1+\sigma}\,\frac{\partial V_z}{\partial z}+\frac{1-2\sigma}{1-\sigma}\,\alpha\,\frac{\lambda}{i \omega\,\rho \, C}\,k_{\perp}^2\,\Theta\right]
\label{B28}
\end{equation}
\begin{equation}
\tilde{\sigma}_{zx}=\tilde{\sigma}_{zy} = \frac{E}{2(1+\sigma)}\left[V_z-\frac{1}{k_{\perp}}\,\frac{\partial V_r}{\partial z}+2\frac{1+\sigma}{1-\sigma}\,\alpha\,\frac{\lambda}{i \omega\,\rho \, C}\,\frac{\partial \Theta}{\partial z}\right] \qquad .
\label{B29}
\end{equation}

Once written a complete solution of the elastic equation, we now consider the boundary conditions. At $z=0$ the null stress condition is
\begin{equation}
\tilde{\sigma}_{zz} \big|_{z=0} =  0
\label{B30}
\end{equation}
\begin{equation}
\tilde{\sigma}_{zx} \big|_{z=0} = \tilde{\sigma}_{zy} \big|_{z=0} = 0  \qquad .
\label{B31}
\end{equation}
The continuity conditions for the stress at the interface between layer and substrate are
\begin{equation}
\tilde{\sigma}_{zz} \big|_{z=d^-} =  \tilde{\sigma}_{zz} \big|_{z=d^+}
\label{B32}
\end{equation}
\begin{equation}
\tilde{\sigma}_{zx} \big|_{z=d^-} =  \tilde{\sigma}_{zx} \big|_{z=d^+} \qquad .
\label{B33}
\end{equation}
The continuity of the displacement at the interface is given by the relation $U_i \big|_{z=d^-} =  U_i \big|_{z=d^+}$, that can be transformed, using Eqs.~(\ref{B15}) and (\ref{B27}), into
\begin{equation}
\left(V_z+\alpha_c \, \frac{1+\sigma_c}{1-\sigma_c}\,\frac{1}{k_c^2}\,\frac{\partial \Theta}{\partial z}\right) \bigg|_{z=d^-} =  \left(V_z+\alpha_s \, \frac{1+\sigma_s}{1-\sigma_s}\,\frac{1}{k_s^2}\,\frac{\partial \Theta}{\partial z}\right) \bigg|_{z=d^+}  
\label{B34}
\end{equation}
\begin{equation}
\left(V_r-k_{\perp}\,\alpha_c \, \frac{1+\sigma_c}{1-\sigma_c}\,\frac{1}{k_c^2}\, \Theta\right) \bigg|_{z=d^-} =  \left(V_r-k_{\perp}\,\alpha_s \, \frac{1+\sigma_s}{1-\sigma_s}\,\frac{1}{k_s^2}\, \Theta\right) \bigg|_{z=d^+}   \qquad .
\label{B35}
\end{equation}
Replacing the solutions (\ref{B9a}-\ref{B9b}) and (\ref{B23}-\ref{B26}) into the Eqs.~(\ref{B30}-\ref{B35}) we obtain a system of 6 linear equations for the 6 variables $V_c, V_s, V_{cc}, V_{ss}, V_a, V_b$, that can be solved using annoying but straightforward algebra. As we will see, for our purpose the meaningful function is just $V_c$. 

\subsection{Measured thermal expansion}

The average displacement measured by the laser beam is given in Eq.~(\ref{ut}). By using the inverse Fourier transform for $u_z(\textbf{r}_{\perp},t)$ we can write
\begin{equation}
\bar{u}(t)=\int \ud \omega \, e^{i\,\omega\,t} \,\delta X (\omega) = \int \ud^2 \mathbf{r}_{\perp}\,\frac{e^{-\frac{r_{\perp}^2}{r_0^2}}}{\pi r_0^2} \int \ud \omega \, e^{i\,\omega\,t} \int \frac{\ud^2 \mathbf{k}_{\perp}}{(2 \pi)^2} \, e^{i\,\mathbf{k}_{\perp}\cdot \mathbf{r}_{\perp}} \, U_z \big|_{z=0}\,  \qquad .
\label{B36}
\end{equation}
and performing the integration over $\textbf{r}_{\perp}$ and over the angular coordinate of $\textbf{k}_{\perp}$, we can express the time-Fourier transformed of such displacement as
\begin{equation}
\delta X (\omega) = \int_0^{\infty} \frac{\ud k_{\perp}}{2 \pi} \, k_{\perp}\,\exp\left(-\frac{k^2_{\perp}\,r_0^2}{4}\right)\,U_z \big|_{z=0}   \qquad .
\label{B37}
\end{equation} 
Using Eqs.~(\ref{B15}), (\ref{B23}) and (\ref{B27}) the function $U_z$ in the integrand of Eq.~(\ref{B37}) is transformed into
\begin{equation}
U_z \big|_{z=0} = V_c\,+\,\alpha_c \, \frac{1+\sigma_c}{1-\sigma_c}\,\frac{1}{k_c^2}\, \frac{\partial\Theta}{\partial z}\bigg|_{z=0} \qquad .
\label{B38}
\end{equation} 
The expression of $U_z \big|_{z=0}$ (and, in particular, of $V_c$) is completely and explicitly determined by the algebra described above. It can be shown that $U_z \big|_{z=0}$ can be written in the following form
\begin{eqnarray}
U_z \big|_{z=0} = \frac{2\alpha_c}{k_c^2} \, 
\Bigg[&\gamma_1(k_{\perp})&\,\left(\frac{\partial \Theta}{\partial z}\bigg|_{z=0}\,+\,k_{\perp}\,\sinh(k_{\perp} d)\,\Theta\big|_{z=d}\,-\,\cosh(k_{\perp} d)\,\frac{\partial \Theta}{\partial z}\bigg|_{z=d^-}\right)\nonumber\\
+\,&\gamma_2(k_{\perp})&\,\left(k_{\perp}\,\Theta\big|_{z=0}\,-\,k_{\perp}\,\cosh(k_{\perp} d)\,\Theta\big|_{z=d}\,+\,\sinh(k_{\perp} d)\,\frac{\partial \Theta}{\partial z}\bigg|_{z=d^-}\right) \Bigg] 
\nonumber\\
+\,\frac{2\alpha_s}{k_s^2} \,&\gamma_3(k_{\perp})&\, \Bigg[\frac{\partial\Theta}{\partial z}\bigg|_{z=d^+}+k_{\perp}\,\Theta\big|_{z=d}\Bigg]  \qquad .
\label{B39}
\end{eqnarray}
The particular combinations of the temperature functions that appear in Eq.~(\ref{B39}) assure the correct frequency dependence for $\omega \rightarrow 0$, compensating the divergence given by $1/k_{c,s}^2$. Therefore, they must be kept as they are. On the other hand, the functions $\gamma_1(k_{\perp})$, $\gamma_2(k_{\perp})$ and $\gamma_3(k_{\perp})$ are rather complex for a practical use. It is useful to give their approximations at the lowest order in $k_{\perp}d$, that read 
\begin{eqnarray}
\gamma_1\,=\,\gamma_1(0)& =& \frac{1}{2}\frac{1+\sigma_c}{1-\sigma_c}\left[1+(1-2\sigma_s)\frac{1+\sigma_s}{1+\sigma_c}\frac{E_c}{E_s}\right]   \nonumber\\
\gamma_2\,=\,\gamma_2(0)& =& \frac{1-\sigma_s^2}{1-\sigma_c}\frac{E_c}{E_s}   \nonumber\\
\gamma_3(0)& = &1+\sigma_s   \qquad .
\label{B39a}
\end{eqnarray}
These three parameters are close to unit, to which they reduce if $\sigma_s = \sigma_c = 0$ and $E_c = E_s$. 
When inserting  $U_z \big|_{z=0}$ in Eq.~(\ref{B37}), the approximated expressions of the $\gamma$'s can be used,  provided that $d << r_0/2$, thanks to the fast decaying Gaussian function of $(k_{\perp}\,r_0/2)$. Replacing $\,\Theta\,$ from Eqs.~(\ref{B9a}-\ref{B9b}) in Eq.~(\ref{B39}), inserting the resulting $U_z \big|_{z=0}$ in Eq.~(\ref{B37}), and after some simplifications, we can write the expected photothermal response to the modulated absorbed power as the sum of two terms, equivalent in the simplified model to the expansions of coating and substrate:
\begin{eqnarray}
\delta X_c \,&=& \, \frac{\alpha_c}{\pi \lambda_c} \,\frac{\omega_c}{i\,\omega}\int_0^\infty \ud \xi\,\,\xi\,e^{-\frac{\xi^2}{2}}\,\delta P_{abs}(\omega) F 
\nonumber\\
&&\bigg[\gamma_1\left(\cosh (\xi\,d/r_0)+\mathcal{R}\frac{\xi}{\xi_c}\sinh (\xi\,d/r_0)-\cosh (\xi_c\,d/r_0)-\mathcal{R}\sinh (\xi_c\,d/r_0)\right)
\nonumber\\
&-&\gamma_2\,\frac{\xi}{\xi_c}\left(\mathcal{R}\cosh (\xi\,d/r_0)+\frac{\xi_c}{\xi}\sinh (\xi\,d/r_0)-\mathcal{R}\cosh (\xi_c\,d/r_0)-\sinh (\xi_c\,d/r_0)\right)\,\bigg]
\label{B41}
\end{eqnarray}
and
\begin{equation}
\delta X_s \,= \, -\frac{\alpha_s\,(1+\sigma_s)}{\pi \lambda_s} \,\frac{\omega_s}{i\,\omega}\int_0^\infty \ud \xi \,F\,\left[1-\frac{\xi}{\xi_s}\right]\,\xi\,e^{-\frac{\xi^2}{2}}\,\delta P_{abs}(\omega)
\label{B42}
\end{equation}
where the adimensional integration variable is $\xi=k_{\perp}\,r_0$, and we have defined to simplify the notation 
\begin{equation}
\xi_{c,s} = r_0\,\,\sqrt{k_{c,s}^2+k_{\perp}^2}=\sqrt{i\frac{\omega}{\omega_{c,s}}+\xi^2}   \qquad .
\label{B43}
\end{equation}
The filter function $F$, derived from Eq.~(\ref{Filtro}) by replacing $k_c \rightarrow \sqrt{k_c^2\,+\,k_{\perp}^2}$, can be written as
\begin{equation}
F = \frac{1}{\cosh[\xi_c\,d/r_0]+\mathcal{R}\sinh[\xi_c\,d/r_0]}  \qquad .
\label{B46}
\end{equation}
In the limit $d = 0$ (no coating layer) the filter $F$ obviously assumes a unit value, and $\delta X_s$ coincides with the results of Refs~\cite{Brag1} and \cite{Cerdonio1} (reported here in Eq.~(\ref{XBRG})). In other words, while not evident at a first glance, it can be shown that
\begin{equation}
\frac{1}{i\,\Omega}\int_0^\infty \ud \xi \,\left[1-\frac{\xi}{\sqrt{i\,\Omega+\xi^2}}\right]\,\xi\,e^{-\frac{\xi^2}{2}}\,=\,G\left(\Omega\right)
\end{equation}
where $G(\Omega)$ is given in Eq.~(\ref{G}).
Concerning the expansion of the coating, it can be shown that  $\delta X_c$ in the high frequency limit reduces to the one-dimensional expression (first term on the right side of Eq.~(\ref{X1dim})), if one set $\gamma_1=\gamma_2=0$.

One further effect to be considered is the thermo-refractive displacement. From Eq.~(\ref{uzthr}), using Eq.~(\ref{B37}) for the average displacement, we can write it as
\begin{equation}
\delta X_{tr} \,=\, \lambda_{vac}\,\beta_{eff}\,\int_0^{\infty} \frac{\ud k_{\perp}}{2 \pi} \, k_{\perp}\,\exp\left(-\frac{k^2_{\perp}\,r_0^2}{4}\right)\,\Theta\big|_{z=0}
\label{B47}
\end{equation}
and replacing $\Theta$ from Eq.~(\ref{B9a})
\begin{equation}
\delta X_{tr} \,=\, \frac{\lambda_{vac}\,\beta_{eff}}{2\pi\,\lambda_c\,r_0}\,\int_0^{\infty} \ud \xi \,\frac{\xi}{\xi_c}\,\frac{\sinh(\xi_c\,d/r_0)+\mathcal{R}\,\cosh (\xi_c\,d/r_0)}{\cosh(\xi_c\,d/r_0)+\mathcal{R}\,\sinh (\xi_c\,d/r_0)}\,e^{-\frac{\xi^2}{2}}\,\delta P_{abs}(\omega)  \qquad .
\label{B48}
\end{equation}
Finally, the total response to the modulated absorbed power is
\begin{equation}
\delta X \,=\, \delta X_c\,+\,\delta X_s\,+\,\delta X_{tr}  \qquad .
\label{Xfinal}
\end{equation}

\newpage

\begin{table}
\label{table1}
\begin{center}
\begin{tabular}{lcccc}
\hline \hline 
 &  SiO$_2$ ($T$ = 300~K) & SiO$_2$ (low $T$) & Si ($T$ = 300~K) & Si ($T$ = 10 K) \\
 \hline \hline 
$C_s$~(J/kg~K) & 790 &$C_{Si0_2} (T)$ &710 &0.4 \\
&Ref. \cite{corning}  & Ref. \cite{zeller} &Ref. \cite{shanks} &Ref. \cite{flubacher} \\
\hline 
$\lambda_s$~(W/m K) & 1.30 &0.12 & 150 &  1600  \\
&Ref. \cite{corning}  & Ref. \cite{zeller,mack}  &Ref. \cite{glassbrenner,shanks}  &Ref. \cite{glassbrenner}  \\
\hline 
$\alpha_s$~($\rm{K^{-1}}$) & 5.2~$10^{-7}$ &$\alpha_{Si0_2} (T)$ &2.6~$10^{-6}$ &4.8~$10^{-10}$ \\
&Ref. \cite{corning}  & Ref. \cite{white,lyon,ackerman} &Ref. \cite{okada} &Ref. \cite{lyon2} \\
\hline 
$\rho_s$~(kg/$\rm{m^3}$) & 2201 & &2330 &\\
&Ref. \cite{corning}  & & & \\
\hline 
$E$~(GPa) & 72 & &130 [100]&\\
&Ref. \cite{corning} &&169 [110]&\\
&&&188 [111]&\\
& & & Ref. \cite{Wortman}& \\
\hline 
$\sigma$ & 0.16 & &0.28 (100)&\\
&Ref. \cite{corning}&&0.21 (110)&\\
&&&0.18 (111)&\\
&  & & Ref. \cite{Wortman}& \\
\hline 
$\frac{\partial n}{\partial T}$~($\rm{K^{-1}}$)&8.57 $10^{-6}$&(8 + 3.3 T) $10^{-8}$\\
&Ref. \cite{Leviton}&Ref. \cite{Leviton} \\
\hline
\hline
\end{tabular}
\caption{Substrate parameters used in the fitting procedure. The expressions for the low-temperature thermal capacitance and expansion coefficient of silica are respectively $\,C_{Si0_2} (T)=(0.64+0.54T-0.021T^2) T^3 \cdot 10^{-3}\,$ and $\,\alpha_{Si0_2} (T)=(2.6-5.2T+2.9T^2-0.91T^3+0.044T^4)\cdot 10^{-9}\,$. The Poisson factors for Si are averaged over the respective planes.}
\end{center}
\end{table}

\newpage

\begin{table}
\label{table2}
\begin{center}
\begin{tabular}{lcc}
\hline \hline 
 &  T = 300~K  & low T  \\
\hline
$\rho_c C_c$~(J/m$^3$~K) & 1.4$\pm$0.3 $\cdot10^6$&1.6$\pm$0.3 $\cdot10^4$\\
$\lambda_c$~(W/m K) & 0.7$\pm$0.3  & 7$\pm$3 $\cdot10^{-2}$  \\
$\alpha_c$~($\rm{K^{-1}}$) & 2.3$\pm$0.3 $\cdot10^{-6}$ &  9$\pm$1.5$\pm$2 $\cdot10^{-8}$  \\
$\beta_{eff}$~($\rm{K^{-1}}$) & 4.4$\pm$0.6 $\cdot10^{-6}$&    \\
$Abs_{SiO2}$ (ppm)& 9.0$\pm$0.5 &  3.8$\pm$0.2  \\
$Abs_{Si}$ (ppm)& 4$\pm$2 &  0.44 $\times Abs_{SiO2}$  \\
$E$~(GPa) & 91$\pm$7 & \\
& Ref. \cite{Crooks} & \\
$\sigma$ & 0.20 & \\
& Ref. \cite{Crooks} &\\
T (K) & & 10.9$\pm$0.3 \\
\hline
\hline
\end{tabular}
\caption{Coating parameters in the fitting procedures. For $\alpha_c$ at low temperature, the first quoted uncertainty reflects the spread in the assumed thermo-optic coefficient of tantala.}
\end{center}
\end{table}

\newpage

\newpage

\newpage

\end{document}